\documentclass[pre,showpacs,preprint,superscriptaddress]{revtex4-1}
\usepackage{amssymb,graphicx,amsbsy}
\usepackage{amsmath}

\begin{document}
\title{Bounds of percolation thresholds on hyperbolic lattices}
\author{Junghoon F. Lee}
\affiliation{School of Computational Sciences, Korea Institute for Advanced
Study, Seoul 130-722, Korea}
\author{Seung Ki Baek}
\affiliation{School of Physics, Korea Institute for Advanced Study, Seoul
130-722, Korea}
\email[Corresponding author, E-mail: ]{seungki@kias.re.kr}

\begin{abstract}
We analytically study bond percolation on hyperbolic lattices obtained by
tiling a hyperbolic plane with constant negative Gaussian curvature. The
quantity of our main concern is $p_{c2}$, the value of occupation
probability where a unique unbounded cluster begins to emerge.
By applying the substitution method to known bounds
of the order-5 pentagonal tiling, we show that $p_{c2} \ge
0.382~508$ for the order-5 square tiling, $p_{c2} \ge 0.472~043$ for its dual,
and $p_{c2} \ge 0.275~768$ for the order-5-4 rhombille tiling.
\end{abstract}


\maketitle

\section{Introduction}

Hyperbolic geometry is an important model of non-Euclidean geometry where
mathematicians have devoted a great deal of efforts since Carl Friedrich
Gauss~\cite{anderson}. In the context of statistical physics, hyperbolic
geometry has served as a conceptual setting to understand geometric
frustrations in glassy materials~\cite{nelson,*auriac,*sausset08}. It also
has a nontrivial connection to the two-dimensional conformal field
theory~\cite{kleban} and there have been attempts to identify it with the
underlying geometry of complex networks~\cite{boguna,*aste}. For this
reason, basic understanding of physical processes on this geometry is
expected to be relevant in a wider context of statistical physics as well.
If we are to discretize a surface by means of regular tiling to study
physical systems defined on a lattice, in particular, hyperbolic geometry
provides infinitely more possibilities than the Euclidean geometry: Let
$\{p,q\}$
denote tiling where $q$ regular $p$-gons meet at each vertex. This bracket
representation is called the Schl\"afli symbol. It is easy to see that a
flat plane admits only three possibilities: $\{3,6\}$ (triangular),
$\{4,4\}$ (square), and $\{6,3\}$ (honeycomb), while every $\{p,q\}$ such
that $(p-2)(q-2) > 4$ describes a hyperbolic plane with constant negative
Gaussian curvature. In other words, each pair of such $\{p,q\}$
defines a hyperbolic lattice that can completely cover the infinite
hyperbolic plane with translational symmetry.
The most important physical property of a hyperbolic plane is that the area
of a circle on it is an exponential function of the radius, which means that the
circumference increases exponentially, too. Therefore, choosing any finite
domain on a hyperbolic lattice, we find that the vertices at the boundary
always occupy a finite portion of the whole number of vertices inside the
domain even if the domain is very large.
This property is called nonamenable in literature~\cite{benjamini}
and makes essential differences in many physical systems from their planar
counterparts.

Percolation is a simple yet most interesting problem of fully geometric
nature, asking the possibility of a global connection out of local
connections~\cite{bollobas}. Let us introduce
the bond percolation problem, which will be studied in this work: For a given
structure of sites and bonds linking them, suppose that each bond is open
with probability $p$ and closed with $1-p$, where the parameter $p$ is called
occupation probability.
One fundamental question in percolation is
to find a critical value $p=p_c$ where an unbounded cluster of open bonds
begins to form.
This question has been already answered for the three regular ways of tiling a
flat plane~\cite{sykes,kesten,honeycomb} and also for more general ones
provided that they allow a generalized cell--dual-cell
transformation~\cite{cell,*exact}. On these flat lattices, there exists a unique
$p_c$ above which the largest cluster occupies a finite fraction of the
system, and the length scale of this cluster becomes unbounded at this point.
On a hyperbolic plane, on the other hand, studies of percolation started
about one decade ago~\cite{benjamini,dual}. The most remarkable prediction
here is that there
generally exist two different percolation thresholds $p_{c1}$ and $p_{c2}$
with $p_{c1} < p_{c2}$, so that an unbounded cluster begins to appear at
$p_{c1}$ while a \emph{unique} unbounded cluster is observed only when $p$
reaches a higher value, $p_{c2}$. Note that this is a consequence of the
nonamenable property~\cite{bollobas} and that these two thresholds coalesce
on a flat plane by $p_{c1} = p_{c2} = p_c$.
Numerical calculations have qualitatively supported this mathematical
prediction~\cite{perc,*ebt}, but a direct numerical estimate of $p_{c2}$ is
usually a difficult task since the system size increases exponentially as
the length scale grows. This has led to a debate about
$p_{c2}$ on some hyperbolic structures~\cite{comment,*reply,*an,*bnd,ziff,cross}.

Recently, nontrivial upper bounds of $p_{c1}$ for self-dual tiling $\{m,m\}$
were derived by a combinatorial argument~\cite{quantum,delfosse}. If $m=5$, for
example, $p_{c1}^{\{5,5\}}$ is bounded as $1/4 \le p_{c1}^{\{5,5\}} \le
0.381~296$. The upper bound is a solution of the following polynomial
equation:
\begin{eqnarray*}
&&-2 + 3p + 3p^2 - 567p^3 + 6721p^4 - 35~655p^5\\
&&+ 115~505p^6 - 257~495p^7 + 41~8210p^8 - 509~100p^9\\
&&+ 469~900p^{10}- 328~480p^{11} + 171~560p^{12} - 65~000p^{13}\\
&&+ 16~900p^{14} - 2700p^{15} + 200p^{16}=0,
\end{eqnarray*}
and the lower bound originates from the simple fact that
\begin{equation}
p_{c1} \ge 1/(n-1)
\label{eq:triv}
\end{equation}
for a lattice $\{m,n\}$ with coordination number $n$.
Gu and Ziff have suggested that $p_{c1}^{\{5,5\}}< 0.346$ and
$p_{c2}^{\{5,5\}}>0.666$,
with estimating $p_{c1}^{\{5,5\}} \approx 0.263$ and $p_{c2}^{\{5,5\}}
\approx 0.749$ by numerically calculating the crossing
probability~\cite{ziff}, and these results also strongly support the above
analytic bounds. From our point of view,
the work in Refs.~\cite{quantum,delfosse} is important in two aspects:
First, it showed the possibility of rigorous analytic bounds free
from any numerical ambiguities. Second, it dealt with the problem from a new
point of view, that is, in terms of the capacity of a quantum erasure
channel.
In this Brief Report, we point out that the nontrivial bounds in
Ref.~\cite{delfosse} also imply nontrivial bounds of other hyperbolic
lattices. Specifically, it is made possible by using the substitution
method~\cite{wierman}, and the lattices considered here will be endowed with
transitivity to remove any undesirable boundary effects, which allows us to
exploit duality properties among the lattices, too. In the next section, we
will briefly explain the substitution method and then show how to apply it
to hyperbolic lattices as well as the results in Sec.~\ref{sec:result}.
This work is summarized in Sec.~\ref{sec:summary}.

\section{Substitution method}
\label{sec:sub}

The easiest way to explain the substitution method is to begin with the
star-triangle transformation~\cite{wierman,may,bollobas}. Then, we proceed
to other cases such as star-square and star-pentagon transformations, which
will be used in our problem. 
Consider a lattice  $\mathcal{L}$ with congruent $n$-gons as its basic
building blocks. By drawing an $n$-star with $n$-bonds inside every $n$-gon,
we obtain another lattice $\mathcal{L}'$, where the occupation probability
is denoted as $q$ in order to avoid confusion with $p$ of $\mathcal{L}$.

\begin{figure}
\includegraphics[width=0.2\textwidth]{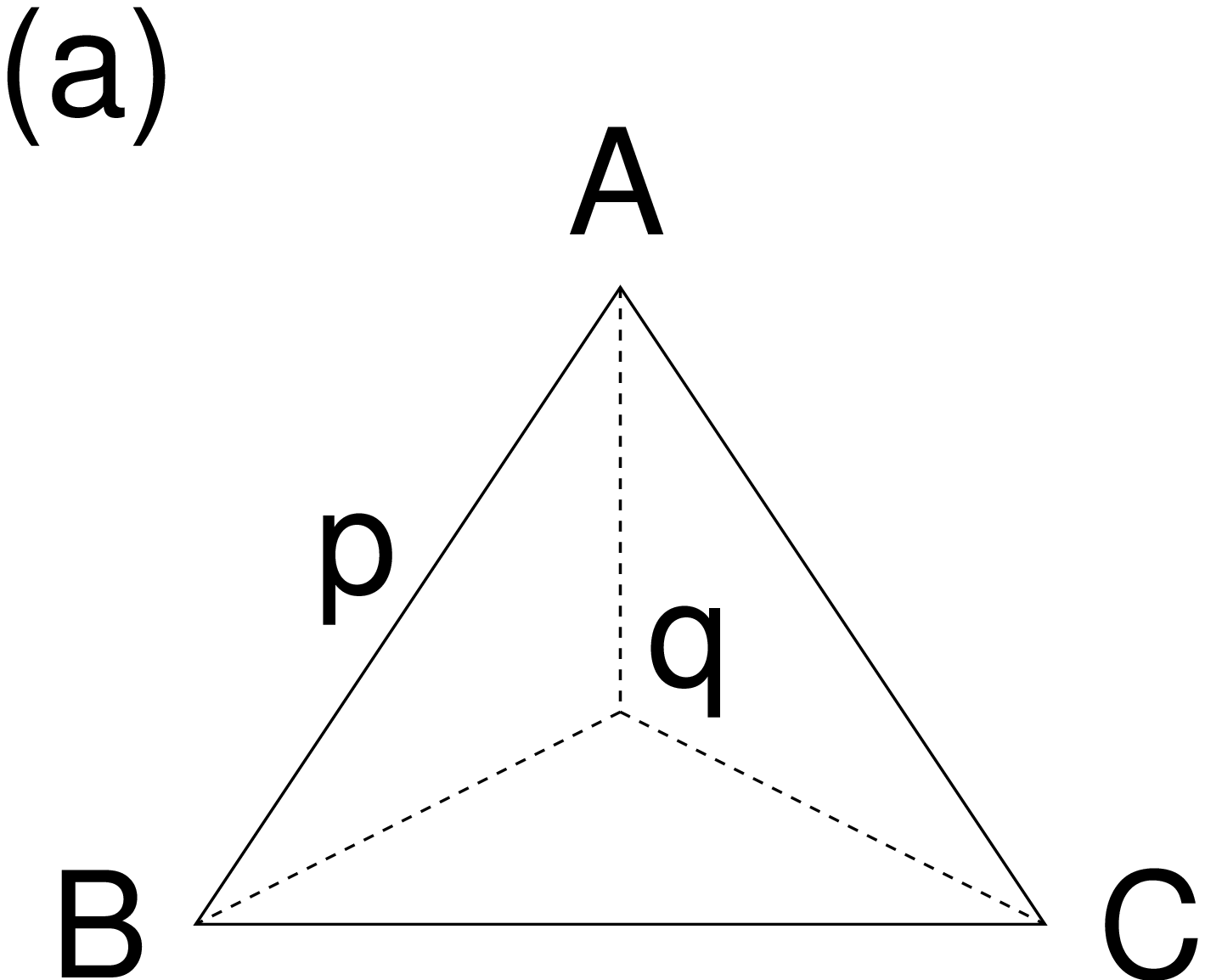}
\includegraphics[width=0.2\textwidth]{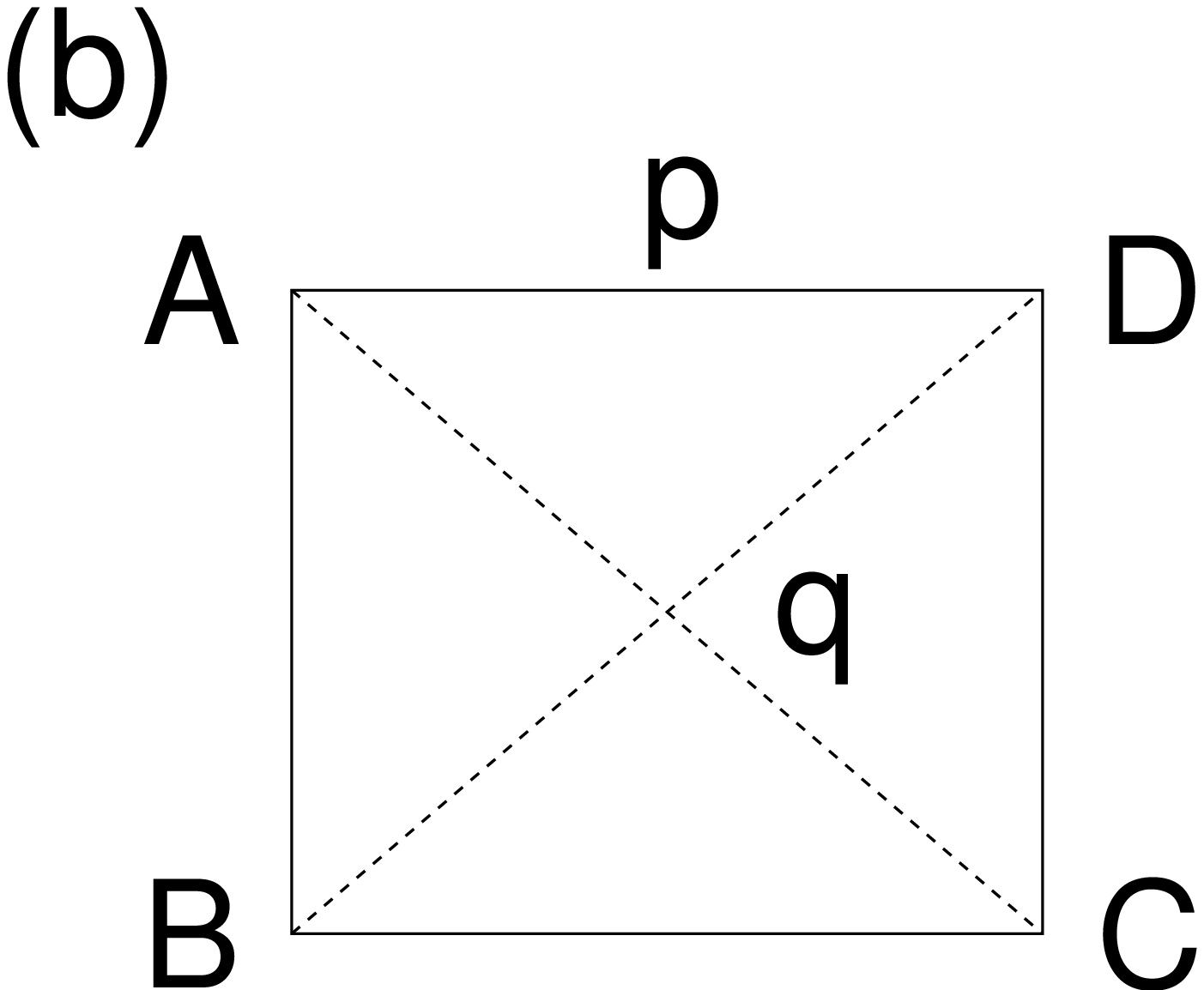}
\includegraphics[width=0.2\textwidth]{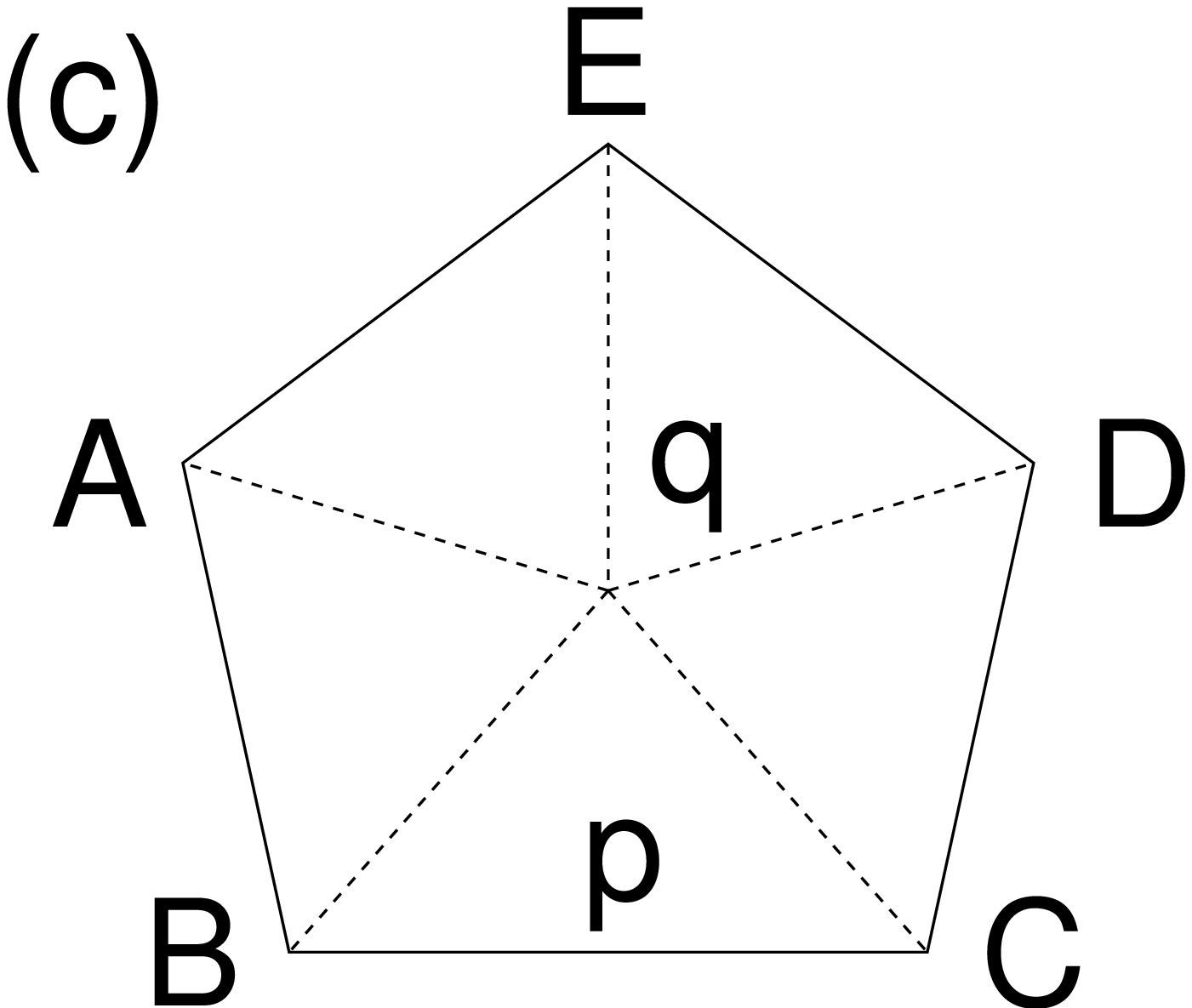}
\caption{Substitution regions of the (a) star-triangle, (b) star-square, and
(c) star-pentagon transformations.
The solid lines are occupied with probability $p$ and the
dashed lines are occupied with $q$.}
\label{fig:sub}
\end{figure}

\subsection{Triangle}

\begin{figure}
\includegraphics[width=0.48\textwidth]{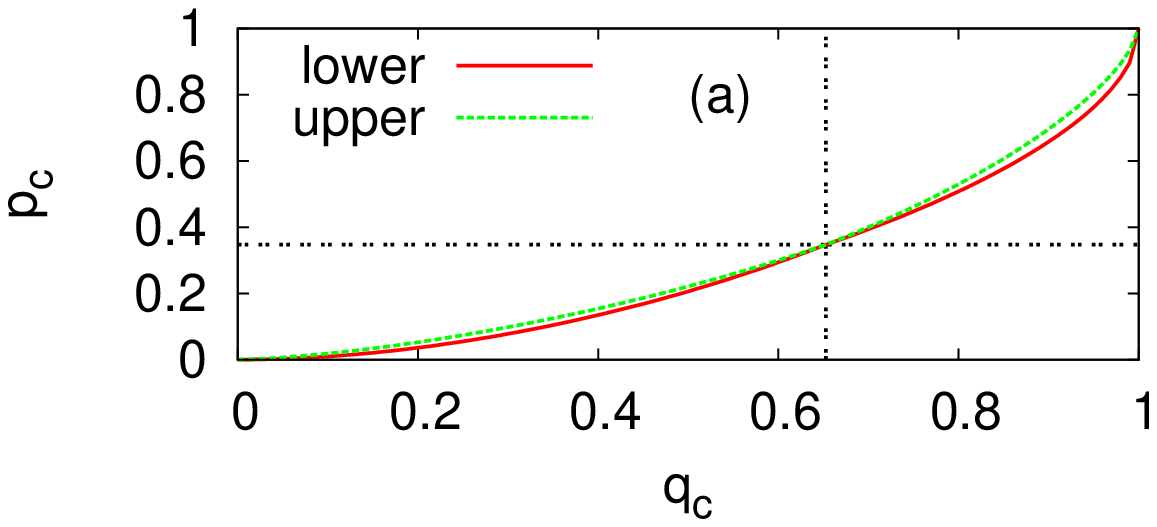}
\includegraphics[width=0.48\textwidth]{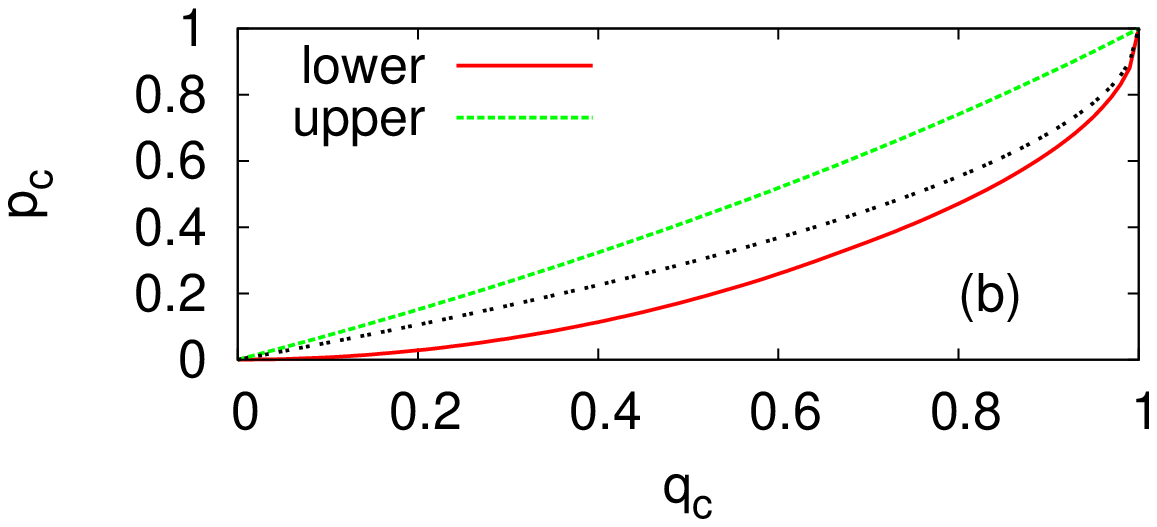}
\includegraphics[width=0.48\textwidth]{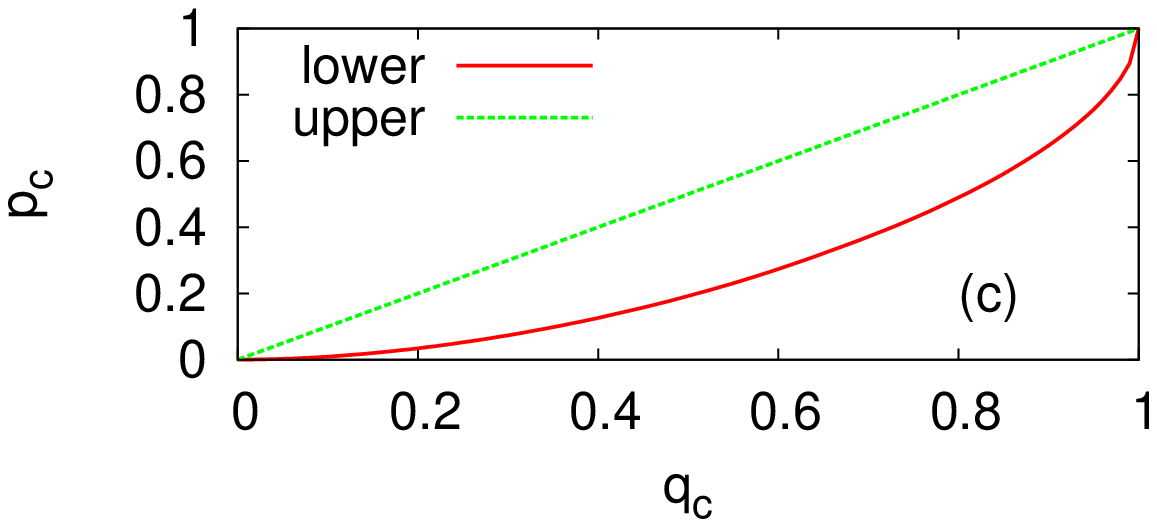}
\caption{(Color online)
Solutions of the inequalities in the (a) star-triangle, (b)
star-square, and (c) star-pentagon transformations, respectively.
In (a), the horizontal dashed line
represents $p_c = 2\sin(\pi/18) \approx 0.347~296$ and the vertical dashed
line represents $q_c = 1-2\sin(\pi/18) \approx 0.652~704$. In (b), the
dashed curve means $p_c = 1-\sqrt{1-q_c}$ to check the square lattice
$\{4,4\}$ (see text).}
\label{fig:bounds}
\end{figure}

If $n=3$, $\mathcal{L}$ is the triangular lattice [Fig.~\ref{fig:sub}(a)],
whereas if $n=4$, $\mathcal{L}$ is the square lattice
[Fig.~\ref{fig:sub}(b)]. To explain the substitution method in a simple
manner, we consider the
$n=3$ case. Suppose that we happen to know the percolation threshold $q_c$
of $\mathcal{L}'$. We wish to find bounds of $p_c$ on $\mathcal{L}$ from the
knowledge of $q_c$. Consider one of the triangles $T$ in $\mathcal{L}$ and its
corresponding star $T'$ in $\mathcal{L}'$ so that $T$ and $T' $ share the
boundary vertices, $A$, $B$, and $C$. We look at all the possible cases
of connection among the boundary vertices $A$, $B$, and $C$
on $T$ and $T'$, respectively. Suppose that $\mathcal{L}$ is in the
supercritical state and $\mathcal{L}'$ is at the critical state
(percolating phase). Then, we can make the following qualitative statement:
it is probable that $T$ has more connectivity among its boundary vertices
than $T'$.
To transform this qualitative statement into a quantitative expression, we
introduce some combinatorial concepts. Let $\mathcal{S}$ be a set of all
the possible partitions of boundary vertices $A$, $B$, and $C$: If $A$ and
$B$ are connected by open bonds and $C$ is separated from them, the
representation is partition $AB|C$. A set of connected vertices in a
partition will be called a block.
For partition $AB|C$, $AB$ and $C$ are two distinct blocks. And for $n$
boundary vertices in general, there are $n$th order Bell numbers of elements
in $\mathcal{S}$~\cite{may}. The set $\mathcal{S}$ can be a
partially ordered set if we define an order: For two partitions $\pi$ and
$\sigma$ of $\mathcal{L}$, we define $\pi \leq \sigma$ if and only if for a
block $b_\pi$ of $\pi$, there exists a block $b_\sigma$ containing $b_\pi$
on $\sigma$. And we say that $\pi$ is more refined than $\sigma$, or that
$\pi$ is a refinement of $\sigma$. For example, $A|B|C \leq AB|C$ and $AB|C
\leq ABC$. But there is no order between $AB|CD$ and $ABC|D$ because there
is no block in $AB|CD$ which covers $ABC$ of $ABC|D$ and vice versa.
A subset $\mathcal{U}$ of $\mathcal{S}$ is an up-set if and only if for
$\pi_1, \pi_2 \in \mathcal{S}$, $\pi_1 \in \mathcal{U}$
and $\pi_1 \leq \pi_2$, then $\pi_2 \in
\mathcal{U}$. By this definition, we can generate nine up-sets as follows:
\begin{eqnarray*}
\mathcal{U}_0 &=& \mathcal{S} \\
\mathcal{U}_1 &=& \{ABC\} \\
\mathcal{U}_2 &=& \{ABC, A|BC\} \\
\mathcal{U}_3 &=& \{ABC, AB|C\} \\
\mathcal{U}_4 &=& \{ABC, B|AC\} \\
\mathcal{U}_5 &=& \{ABC, A|BC, AB|C\} \\
\mathcal{U}_6 &=& \{ABC, AB|C, B|AC\} \\
\mathcal{U}_7 &=& \{ABC, A|BC, B|AC\} \\
\mathcal{U}_8 &=& \{ABC, A|BC, AB|C,B|AC\}
\end{eqnarray*}
Let $P_p (\mathcal{U})$ and $Q_q(\mathcal{U})$ be probabilities that $T$
and $T'$ form an element partition of up-set $\mathcal{U}$, respectively.
Then, we rewrite the qualitative statement as such $P_p (\mathcal{U}) \geq
Q_{q_c} (\mathcal{U})$ for every up-set $\mathcal{U}$ of $\mathcal{S}$ with
probability $1$. 
We can solve this inequality with respect to $p$ and get a solid
interval of $p_m\leq p \leq1$, for $P_p (\mathcal{U})$ is known as a monotone
increasing polynomial function of $p$
with $P_{p=0}(\mathcal{U})=0$ and $P_{p=1}(\mathcal{U})=1$ for every up-set
$\mathcal{U} \neq \mathcal{S}$ while 
$Q_{q_c} (\mathcal{U})$ is a constant. Here, $p_m$ is a lower bound of the
percolation threshold of $\mathcal{L}$ since if $p$ is smaller than $p_m$,
the above inequality cannot hold and $\mathcal{L}$ cannot be in the
supercritical state. On the contrary, suppose that $\mathcal{L}$ is in the
subcritical state and $\mathcal{L}'$ is at the critical state. Similarly, we
can conclude $P_p (\mathcal{U}) \leq Q_{q_c} (\mathcal{U})$. By solving
this inequality again, we get another interval, $0 \leq p \leq p_M$, where
$p_M$ is an upper bound of the percolation threshold of
$\mathcal{L}$. By taking the intersection of the two intervals, we obtain $p_m
\leq p \leq p_M$. This is the interval in which the percolation threshold of
$\mathcal{L}$ can
exist. In $n=3$ case, the resulting set of inequalities with respect to all
its up-sets is found as
$P_\mathcal{L} [\mathcal{U}_i] \leq P_{\mathcal{L}' } [\mathcal{U}_i]$
with $i=0,\ldots,8$.
In fact, by symmetry and triviality, seven of them turn out to be redundant
or trivial so we are left with
\begin{eqnarray*}
&3p^2(1-p) + p^3 \leq q_c^3,\\
&3p(1-p)^2 + 3p^2(1-p) +p^3 \leq 3q_c^2(1-q_c) + q_c^3,
\end{eqnarray*}
with $0 \leq p \leq1$.
The largest value satisfying all these inequalities for given $q_c$ is a
lower bound of $p_c$. In order to find an upper bound for $p_c$,
we need to revert both the inequalities above and the smallest $p$ satisfying
the reversed inequalities gives us an upper bound of $p_c$.
The results are shown in Fig.~\ref{fig:bounds}(a). There is one point where
the upper and lower bounds coalesce, that is, $(q_c, p_c)=\left(1-2\sin
\frac{\pi}{18}, 2\sin \frac{\pi}{18}\right)$. Here, the star-triangle
transformation yields the exact percolation threshold $p_c = 2\sin(\pi/18)$
for the triangular lattice $\{3,6\}$ and $q_c = 1- 2\sin(\pi/18)$ for the
honeycomb lattice $\{6,3\}$~\cite{honeycomb}.
But generally, the substitution method gives us an interval in which the
percolation threshold can exist.


\subsection{Square}

Let us now turn our attention to the star-square case shown in
Fig.~\ref{fig:bounds}(b). By enumerating all the possible $345$ up-sets, we
find $53$ different inequalities. Many of them are redundant, however, and
we need to consider only the following inequalities:
\begin{eqnarray*}
&4p^2 - 4p^3 + p^4 \leq 2q_c^2 - q_c^4,\\
&2p^2 - p^4 \leq q_c^4,\\
&4p - 6p^2 + 4p^3 - p^4 \leq 4q_c^2 - 4q_c^3 + q_c^4,
\end{eqnarray*}
with $0<p<1$ in order to find a lower bound of $p_c$ for given $q_c$. The
first inequality is for an up-set $\{ABCD$, $A|BCD$, $B|ACD$, $C|ABD$,
$D|ABC$, $AC|BD$, $A|C|BD$, $B|D|AC\}$,
generated by $\{A|C|BD$, $B|D|AC\}$.
The second inequality is for $\{ABCD$, $AB|CD$, $AD|BC\}$ generated by
$\{AB|CD$, $AD|BC\}$, and the third
is for an up-set generated by $\{A|B|CD$, $B|C|AD$, $C|D|AB$, $A|D|BC\}$,
respectively. As above, reverting all the three inequalities will yield an
upper bound for $p_c$, and
the results are plotted in Fig.~\ref{fig:bounds}(b).

It is interesting to
consider the square lattice $\{4,4\}$ since the star-square transformation
transforms a square lattice with double bonds to another square lattice,
rotated by angle $\pi/4$ from the original lattice. Then, $p_c$ and $q_c$
should be related by $p_c = 1-\sqrt{1-q_c}$. Our upper and lower bounds
include this relationship over the whole region of $q_c$. If this happened
only within a limited region of $q_c$, we could obtain nontrivial bounds for
$p_c$ directly from this plot.
Although this is not the case, this example shows
that the substitution method indeed yields correct results.

\subsection{Pentagon}

For the star-pentagon case, the number of possible up-sets is $161~166$,
from which $1237$ different inequalities are found. Once again, most of them
are redundant, and the set of inequalities to solve turns out to be
\begin{eqnarray*}
&5p - 10p^2 + 10p^3 - 5p^4 + p^5 \leq 5q^2 - 5q^3 + q^5,\\
&5p^3 - 5p^4 + p^5 \leq q^5,\\
&5p^2 - 5p^3 + p^5 \leq 5q_c^2 - 5q_c^3 + q_c^5,
\end{eqnarray*}
with $0\leq p \leq 1$ when we are to find a lower bound.
The first inequality is for an up-set generated by $\{AB|C|D|E$, $AE|B|C|D$,
$BC|A|D|E$, $CD|A|B|E$, $DE|A|B|C\}$.
The second inequality is for an up-set generated by $\{AB|CDE$, $AE|BCD$,
$BC|ADE$, $CD|ABE$, $DE|ABC\}$. Finally, the third inequality is for an
up-set generated by $\{A|B|CDE$, $A|E|BCD$, $B|C|ADE$, $C|D|ABE$,
$D|E|ABC\}$ and essentially the same as $p\leq q_c$.
Finding an upper bound is also straightforward. The results are shown in
Fig.~\ref{fig:bounds}(c).

\section{Results}
\label{sec:result}

\begin{figure}
\includegraphics[width=0.25\textwidth]{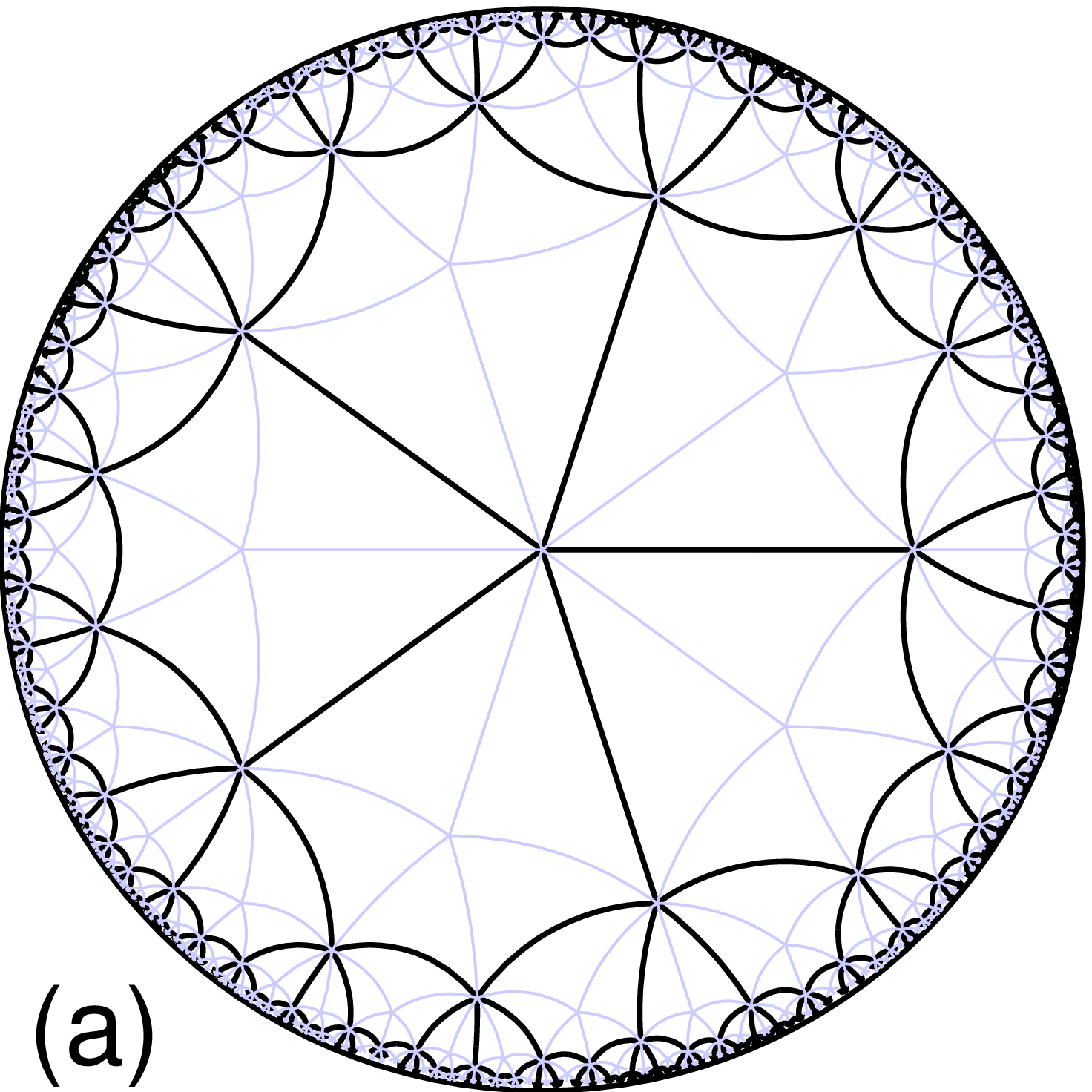}
\includegraphics[width=0.25\textwidth]{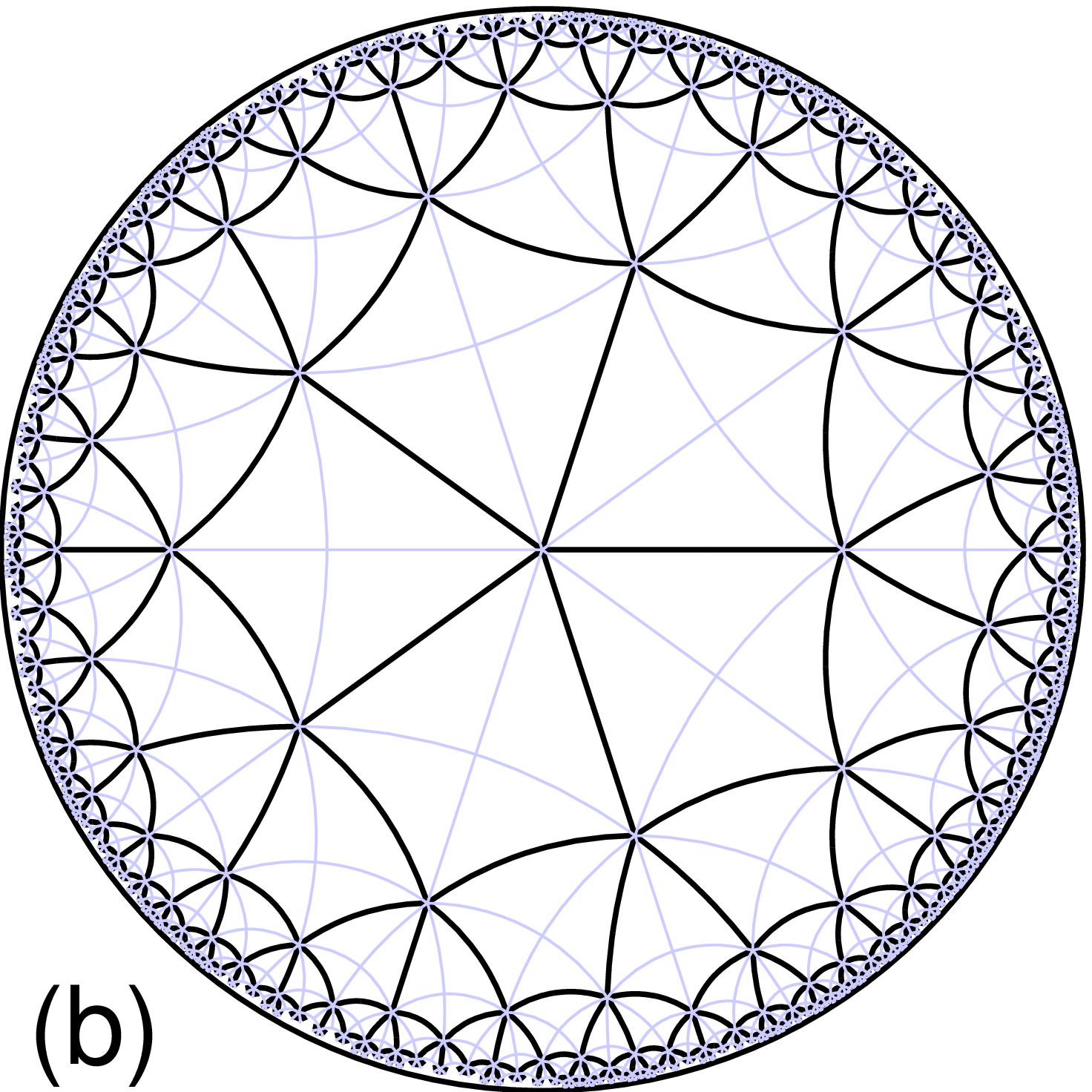}
\includegraphics[width=0.25\textwidth]{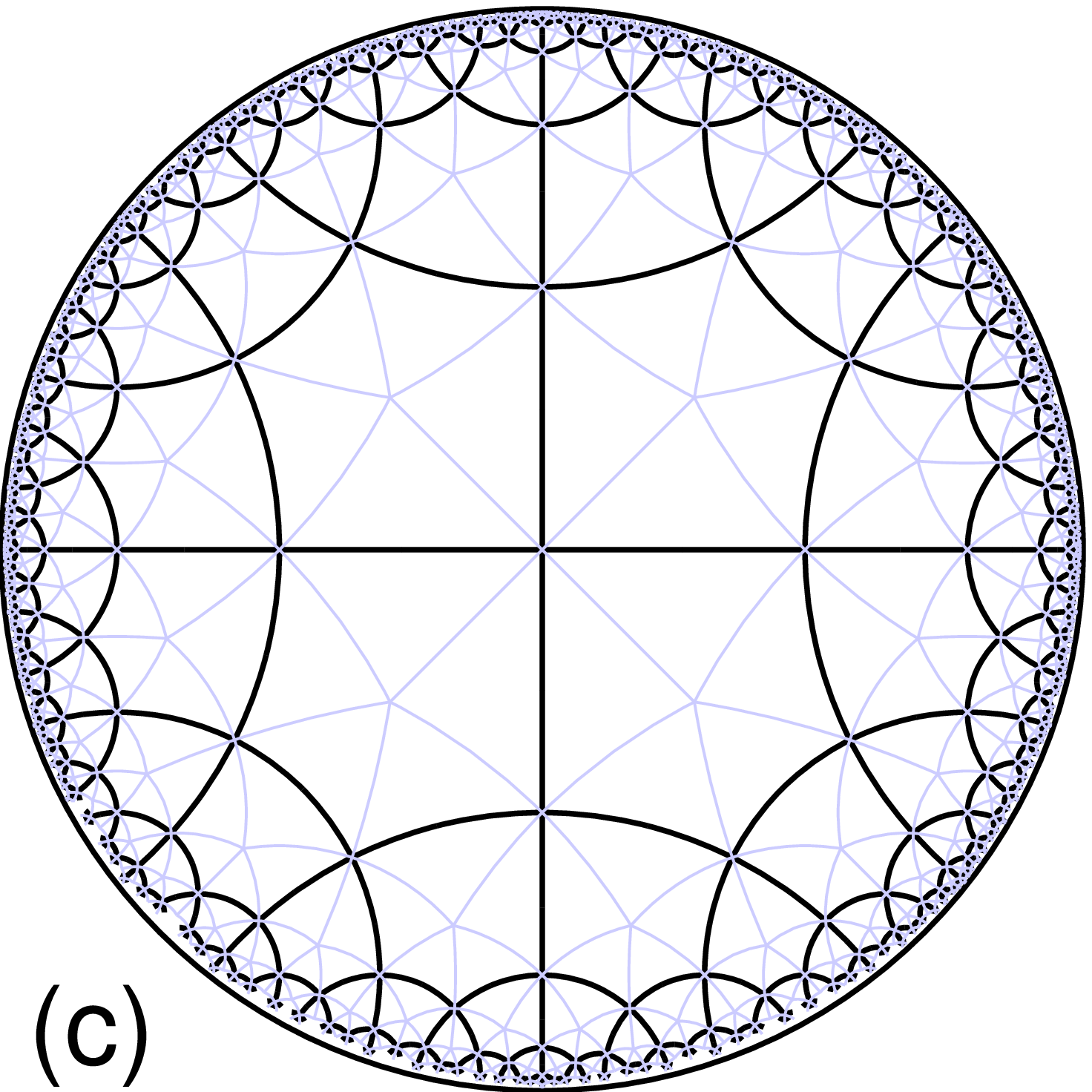}
\caption{(Color online)
Lattice structures depicted on the Poincar\'e disk~\cite{[{The
figures are generated by a java applet developed by }][{}]plunk}.
(a) The star-pentagon transformation of the order-5 pentagonal
tiling $\{5,5\}$ (black) leads to the order-5 square tiling $\{4,5\}$
(gray). (b) The star-square transformation of $\{4,5\}$ (black) leads to the
order-5-4 rhombille tiling (gray). (c) The star-pentagon transformation
relates the order-4 pentagonal tiling $\{5,4\}$ (black) to the order-5-4
rhombille tiling (gray). If shifted by one lattice spacing, the order-5-4
rhombille tiling in (c) looks the same as that in (b).
}
\label{fig:lattice}
\end{figure}

\subsection{Order-5 square tiling}

By applying the star-pentagon transformation to the order-5 pentagonal
tiling $\{5,5\}$ with double bonds, we find the order-5 square tiling
$\{4,5\}$ [Fig.~\ref{fig:lattice}(a)]. Note that we need double bonds
in order to distribute five bonds to every pentagonal face.
Recall that the threshold $p_{c1}$ of $\{5,5\}$ is bounded as $1/4 \leq
p_{c1}^{\{5,5\}} \leq 0.381~296$~\cite{quantum}, which automatically implies
$0.618~704 \leq p_{c2}^{\{5,5\}} \leq 3/4$ by self-duality since the duality
implies
\begin{equation}
\begin{array}{l}
p_{c1}^{\{m,n\}} + p_{c2}^{\{n,m\}} = 1,\\
p_{c2}^{\{m,n\}} + p_{c1}^{\{n,m\}} = 1,
\end{array}
\label{eq:duality}
\end{equation}
for lattices represented by $\{m,n\}$ and $\{n,m\}$~\cite{dual}.
If every neighboring pair of vertices in $\{5,5\}$ are
connected by double bonds with occupation probability $p'$, the
corresponding bounds of the critical threshold are located by the simple
relation $p_c' = 1-\sqrt{1-p_c}$ as
\begin{eqnarray*}
&&0.133~975 \leq {p'}_{c1}^{\{5,5\}} \leq 0.213~423,\\
&&0.382~508 \leq {p'}_{c2}^{\{5,5\}} \leq 0.5.
\end{eqnarray*}
Then, solving the inequalities for the star-pentagon case, we obtain bounds
for $\{4,5\}$. The detailed procedure is given as follows: Suppose that
${p'}_{c1}^{\{5,5\}} = 0.133~975$. The corresponding bounds are $0.133~975
\le p_{c1}^{\{4,5\}} \le 0.413~131$, whereas if ${p'}_{c1}^{\{5,5\}} =
0.213~423$, the bounds are $0.213~423 \le p_{c1}^{\{4,5\}} \le 0.527~957$.
Therefore, the resulting
bounds should be $0.133~975 \leq p_{c1}^{\{4,5\}} \leq 0.527~957$ in total,
and the same
reasoning yields $0.382~508 \leq p_{c2}^{\{4,5\}} \leq 0.807~697$.
However, Eq.~(\ref{eq:triv})
further constrains $p_{c1}^{\{4,5\}}$ as larger than or equal to $1/4$.
Likewise, we see that $p_{c1}^{\{5,4\}} \ge 1/3$, which implies
$p_{c2}^{\{5,4\}} \leq 2/3$ from the duality [Eq.~(\ref{eq:duality})].
We thus conclude that
\begin{eqnarray*}
&&1/4 \leq p_{c1}^{\{4,5\}} \leq 0.527~957,\\
&&0.382~508 \le p_{c2}^{\{4,5\}} \leq 2/3.
\end{eqnarray*}

\subsection{Order-5-4 rhombille tiling}

The same procedure can be repeated on the order-5 square tiling $\{4,5\}$.
The star-square transformation changes it to the order-5 rhombille
tiling, whose face configuration can be denoted by $V4.5.4.5$
[Fig.~\ref{fig:lattice}(b)]. The face configuration means the numbers of
faces at each of vertices around a face.
The computation is similar to the
above one: By putting double bonds between every pair of vertices in
$\{4,5\}$, we see that the critical probabilities are bounded as
\begin{eqnarray*}
&&0.133~975 \le {p'}_{c1}^{\{4,5\}} \le 0.312~946,\\
&&0.214~194 \le {p'}_{c2}^{\{4,5\}} \le 0.42265.
\end{eqnarray*}
When the star-square transformation is applied, it is straightforward to obtain
\begin{eqnarray*}
&&0.178~197 \le p_{c1}^{V5.4.5.4} \le 0.656~963,\\
&&0.275~768 \le p_{c2}^{V5.4.5.4} \le 0.760~854,
\end{eqnarray*}
but some are no better than trivial since
$p_{c1}^{V5.4.5.4} \ge 1/4$ for coordination number $n \le 5$, and
the dual of $V5.4.5.4$, called the tetrapentagonal tiling, has $p_{c1}\ge 1/3$
with coordination number $4$. The bounds
for this order-5-4 rhombille tiling are therefore found to be
\begin{eqnarray*}
&&1/4 \le p_{c1}^{V5.4.5.4} \le 0.656~963,\\
&&0.275~768 \le p_{c2}^{V5.4.5.4} \le 2/3.
\end{eqnarray*}

\subsection{Order-4 pentagonal tiling}

It is notable that the order-4 pentagonal tiling
$\{5,4\}$ is also related to the order-5-4 rhombille tiling by the
star-pentagon transformation [Fig.~\ref{fig:lattice}(c)]. Since the duality
[Eq.~(\ref{eq:duality})] also imposes conditions for thresholds in $\{5,4\}$
and $\{4,5\}$ as
\begin{eqnarray*}
&&p_{c1}^{\{4,5\}} + p_{c2}^{\{5,4\}} = 1,\\
&&p_{c2}^{\{4,5\}} + p_{c1}^{\{5,4\}} = 1,
\end{eqnarray*}
one could expect sharper bounds by exploiting both the relations, i.e., the
star-pentagon transformation and the duality transformation. Unfortunately,
since the star-pentagon transformation yields too large bounds
[see Fig.~\ref{fig:bounds}(c)], it adds no information to the
duality results, which are expressed as
\begin{eqnarray*}
&&1/3 \le p_{c1}^{\{5,4\}} \le 0.617~492,\\
&&0.472~043 \le p_{c2}^{\{5,4\}} \le 3/4,
\end{eqnarray*}
where $1/3$ is a trivial bound from the coordination number $n=4$ and
$p_{c2}^{\{5,4\}} \le 3/4$ is a direct consequence of $p_{c1}^{\{4,5\}} \ge
1/4$.

\section{Summary}
\label{sec:summary}

\begin{table}
\caption{Analytic bounds of bond percolation thresholds on hyperbolic lattices.}
\begin{tabular*}{\hsize}{@{\extracolsep{\fill}}cccc}\\ \hline \hline
tiling & lower threshold & upper threshold & method \\ \hline
order-5 pentagon & $1/4 \le p_{c1} \le 0.381~296$ & $0.618~704 \le p_{c2}
\le 3/4$ & Ref.~\cite{delfosse}\\
order-5 square & $1/4 \le p_{c1} \le 0.527~957$ & $0.382~508 \le p_{c2} \le
2/3$ & substitution \\
order-4 pentagon & $1/3 \le p_{c1} \le 0.617~492$ & $0.472~043 \le p_{c2}
\le 3/4$ & duality \\
order-5-4 rhombille & $1/4 \le p_{c1} \le 0.656~963$ & $0.275~768 \le p_{c2}
\le 2/3$ & substitution \\
\hline \hline
\end{tabular*}
\label{table:summary}
\end{table}

In summary, we have obtained analytic bounds of percolation thresholds on
three hyperbolic lattices by applying the substitution method to the known
bounds for the order-5 pentagonal tiling $\{5,5\}$.
Our results are summarized in Table~\ref{table:summary}. The obtained bounds
are admittedly too broad to be very informative. But our approach
illustrates how analytic bounds for one lattice can be made useful in
estimating those for other lattices tiling a hyperbolic plane.
Precise knowledge of $p_{c2}$ is still greatly needed in studies of
percolation on hyperbolic lattices in general, and we hope that this
approach can make further progress in future studies.

\acknowledgments
S.K.B. thanks R. M. Ziff for having informed him about Ref.~\cite{delfosse}.
We are grateful to C. Hyeon for helpful discussions and support.

%
\end{document}